\documentclass[twocolumn,showpacs,amssymb,aps]{revtex4}
\usepackage{amsmath}
\usepackage{amssymb}
\usepackage{graphicx}

\def\be{\begin{equation}}
\def\ee{\end{equation}}
\def\bea{\begin{eqnarray}}
\def\eea{\end{eqnarray}}
\def\ba{\begin{array}}
\def\ea{\end{array}}
\def\ben{\begin{enumerate}}
\def\een{\end{enumerate}}

\def\nnu{\nonumber}

\begin{document}
\newcommand{\half}{{\textstyle\frac{1}{2}}}
\newcommand{\eqn}[1]{(\ref{#1})}
\allowdisplaybreaks[3]

\def\a{\alpha}
\def\b{\beta}
\def\g{\gamma}\def\G{\Gamma}
\def\d{\delta}\def\D{\Delta}
\def\ep{\epsilon}
\def\et{\eta}
\def\z{\zeta}
\def\t{\theta}\def\T{\Theta}
\def\l{\lambda}\def\L{\Lambda}
\def\m{\mu}
\def\f{\phi}\def\F{\Phi}
\def\n{\nu}
\def\p{\psi}\def\P{\Psi}
\def\r{\rho}
\def\s{\sigma}\def\S{\Sigma}
\def\ta{\tau}
\def\x{\chi}
\def\o{\omega}\def\O{\Omega}
\def\k{\kappa}
\def\pa {\partial}
\def\ov{\over}
\def\br{\\}
\def\ud{\underline}
%
\bigskip\bigskip

\title{Origin of the geometric tachyon}

\author{ Ashok Das$^{a,c}$, 
Sudhakar Panda$^b$ and
Shibaji Roy$^c$\footnote{e-mails: das@pas.rochester.edu, panda@mri.ernet.in, shibaji.roy@saha.ac.in}}

\affiliation{$^a$ Department of Physics and Astronomy, University of Rochester, Rochester, 
NY 14627, USA}
\affiliation{$^b$ Harish-Chandra Research Institute,
Chhatnag Road, Jhusi, Allahabad 211019, India}
\affiliation{$^c$ Saha Institute of Nuclear Physics,
1/AF Bidhannagar, Calcutta-700 064, India} 

\bigskip
\begin{abstract}
  
The motion of a D$p$-brane in the background of a stack of coincident NS5-branes
is analyzed as the motion of a relativistic point particle in the transverse space of the
five-branes. In this system, the particle experiences a proper acceleration 
orthogonal to its  proper velocity due to the background dilaton field which 
changes the dynamics from that of a simple geodesic motion. In particular, we
show that in the vicinity of the five-branes, it is this acceleration 
which is responsible for modifying the motion of the radial mode to that of an 
inverted simple harmonic oscillator leading to the tachyonic instability.

\end{abstract} 

\maketitle

In string theory, it is well-known that soliton solutions like the NS5-brane and 
D$p$-brane are contained in the string spectrum. While the former is
supersymmetric and stable, the latter can be either supersymmetric and stable
(BPS D$p$-brane) or non-supersymmetric and unstable (non-BPS D$p$-brane). A pair of  
D-${\bar {\rm D}}$ branes  is non-supersymmetric even if  each
is supersymmetric individually. Such non-supersymmetric systems are unstable because the
lowest lying state of the open string, either (both ends) ending on a single non-BPS D-brane
or stretched between the brane and the antibrane, is a tachyon. The
condensation of the tachyon can lead to a stable brane configuration or the
total annihilation of the brane. A nice review of this phenomenon can be found in \cite{Sen:2004nf}.
The dynamics of tachyon condensation has led to 
many interesting time dependent phenomena including the decay of a non-BPS
brane into a strange ``tachyon matter'' state whose equation
of state is that of a pressureless fluid \cite{Sen:2002in}. An effective action of
the Dirac-Born-Infeld (DBI) type for the tachyon 
\cite{Sen:1999md,Garousi:2000tr,Bergshoeff:2000dq,Kluson:2000iy} has been 
very useful in understanding such processes. 

A completely different dynamics, namely, that of a BPS D$p$-brane
propagating in the background of a stack of coincident NS5-branes, has been
studied recently using the effective DBI action \cite{Kutasov:2004dj}. It 
has been observed there that
when the D$p$-brane comes close to the NS5-branes, the dynamics of the D$p$-brane 
can be mapped to that of the open string tachyon condensation, where the radial mode
on  the D$p$-brane plays the role of the tachyon. Moreover, its equation of
state approaches that of a pressureless fluid with the pressure falling off
exponentially at late times. These properties  along with the fact that a parallel
D$p$-NS5 brane system is a non-supersymmetric system, has led to the
nomenclature ``geometric tachyon" for the radial mode on the D$p$-brane. The tachyonic instability
has been made more precise \cite{Kutasov:2004ct} by compactifying one of the transverse
directions of the NS5-branes on a circle and placing the D$p$-brane, as a
point on this circle diametrically opposite to the five-branes. In such a case it has been
observed that the potential energy density of the D$p$-brane at
this point has a saddle point. As a result, this point corresponds to an
unstable equilibrium and the D$p$-brane develops a tachyonic mode associated with
translations along the circle. Various other aspects of this
system have also been investigated in 
\cite{Kluson:2005qx,Chen:2004vw,Chen:2005wm,Thomas:2005fw,Thomas:2005fu,Panda:2005sg,Panigrahi:2007sq}. 
For example, it has been noted recently  
that under certain conditions a geometric tachyon in one system gets mapped
to the universal open string tachyon in another system and {\em vice-versa} 
\cite{Sen:2007cz}.

In this letter we look for a better understanding of the origin of the tachyonic instability
when the motion is in the uncompactified transverse space. Our interest is in the dynamics of a BPS 
D$p$-brane in type II string
theory in the presence of $N$ coincident NS5-branes and we formulate 
this as the motion of a relativistic point particle in the background of
fields generated by the NS5 branes. We take the five-branes to
be stretched in the directions ($x^1,x^2,\ldots,x^5$) and their 
world-volume directions are denoted by $x^{\bar{\mu}}$, $\bar{\mu}=0,1,\ldots,5$. Similarly
the transverse directions are labeled by $x^m$, $m=6,7,8,9$. The D$p$-brane is
taken to be parallel to the coincident five-branes, i.e., it is extended in
the directions $x^\mu$, $\mu=0,1,\ldots,p$ with $p\leq 5$. Hence the
D$p$-brane is point-like in the directions $x^m$. When the D$p$-brane is
placed at a large distance $r=(x^mx^m)^{1/2}$, from the stack of
coincident NS5-branes, it experiences an attractive force due to both gravitational
and dilatonic interactions. Since the mass ($\sim 1/g_s^2$ where $g_s$ denotes the
string coupling constant) of the NS5-branes is much larger than the mass
($\sim 1/g_s$) of the D$p$-brane at weak string coupling, the D$p$-brane
will move towards the NS5-branes.

The background fields for the five-branes are obtained from the supergravity
solution as ($\eta_{\bar \mu \bar \nu} = (-,+,\cdots,+)$)
\bea
ds^2 &=& - d\tau^{2} = \eta_{\bar \mu \bar \nu} dx^{\bar \mu} dx^{\bar \nu} + G_{mn} dx^m dx^n,\nnu\\
G_{mn} & = & H(r)\delta_{mn},\nnu\\
e^{2(\phi-\phi_0)} &=& H(r) = 1 + \frac{N \ell_s^2}{r^2},\nnu\\
H_{mnp} &=& - \epsilon^q_{mnp} \partial_q \phi,\label{backgrounds}
\eea
where $H(r)$ is the harmonic
function describing the $N$ coincident five-branes (with $\ell_{s}$ 
denoting the fundamental string length). Here $\phi$ denotes the dilaton
field, $e^{\phi_0}= g_s$ is the string coupling constant and $H_{mnp}$ is the
field-strength of the NS $B$-field.

We label the world-volume coordinates of the D$p$-brane by $\xi^\mu$,
$\mu=0,1,\ldots,p$ and identify $\xi^\mu = x^\mu$ by making use of the
reparametrization invariance on the world-volume. Since, we are considering
the motion of the D$p$-brane only in the transverse space of the NS5-branes,
the position of the D$p$-brane in this space, ($x^6, x^7, x^8, x^9$), gives
rise to scalar fields, $X^m(\xi^\mu)$, $m=6,\ldots,9$ on the world-volume of
the D$p$-brane. The dynamics of these scalar fields is governed by the DBI action 
\be
S_p = -\tau_p\int d^{p+1} \xi e^{-(\phi-\phi_0)} \sqrt{-{\rm
    det}\left(G_{\mu\nu}+ B_{\mu\nu}\right)}, \label{dbiaction}
\ee
where $\tau_p$ is the tension of the D$p$-brane. $G_{\mu\nu}$ and $B_{\mu\nu}$
are the induced metric and the $B$-field on the world-volume, given by ($X^{\bar{\mu}} = x^{\bar{\mu}}$)
\bea
G_{\mu\nu} &=& \frac{\partial X^A}{\partial\xi^\mu} \frac{\partial
  X^B}{\partial\xi^\nu} G_{AB}(X),\nnu\\
B_{\mu\nu} &=& \frac{\partial X^A}{\partial\xi^\mu} \frac{\partial
  X^B}{\partial\xi^\nu} B_{AB}(X).\label{inducedmetric}
\eea
Here the indices $A,B=0,1,\ldots,9$ and $G_{AB}$ and $B_{AB}$ are the metric
and the $B$-field in ten dimensions, given in \eqref{backgrounds}. We are interested in
the case when the fields representing the position of the D$p$-brane $X^m$,
$m=6,\ldots,9$, depend only on time, $X^m = X^m(t)$. In this case, the action
\eqref{dbiaction} simplifies considerably and takes the form
\be
S_p = -\tau_p V_p\int dt\, e^{-(\phi-\phi_0)} \sqrt{1- G_{mn} 
\dot X^m \dot X^n},\label{particleaction}
\ee
where $V_p$ is the volume of the $p$-dimensional space in which the D$p$-brane
is stretched out and an `overdot' represents a derivative with respect to
$t$. The dilaton field and the metric $G_{mn}$ are related to the harmonic
function as noted in \eqref{backgrounds}.

We rewrite the above action in a suggestive form as,
\bea
S_p &=& - \tau_p V_p \int dt e^{-\bar{\phi}} \sqrt{-G_{\bar{m}\bar{n}} 
\dot X^{\bar{m}} \dot X^{\bar{n}}}\nnu\\
&=& -\tau_p V_p \int d\tau e^{-\bar{\phi}} \sqrt{-G_{\bar{m}\bar{n}} \frac{dX^{\bar{m}}}{d\tau}
\frac{dX^{\bar{n}}}{d\tau}},\label{covariantparticleaction}
\eea
where $\bar{\phi} = \phi-\phi_0$, $G_{\bar{m}\bar{n}} = (-1,\, G_{mn})$ with
$\bar{m}$ and $\bar{n}$ taking values $\bar{m} = (0, m) =0,6,7,8,9$; 
$X^0 = t$ and $\tau$ is the proper time.  The action \eqref{covariantparticleaction} can be
thought of as describing the dynamics of a relativistic point particle in gravitational as well
as dilatonic backgrounds. Note that $\bar{\phi}$ does not depend on 
$X^0$ and, therefore,  on time explicitly. However, since it depends upon $X^m$, which is a
function of time, it has an implicit time dependence. The Lorentz factor follows from \eqref{backgrounds} to be 
\be
\gamma = \frac{dt}{d\tau} = \frac{1}{\sqrt{-G_{\bar{m}\bar{n}} \dot X^{\bar{m}} \dot X^{\bar{n}}}},\label{lorentzfactor}
\ee
and denoting the proper velocity of the particle as $u^{\bar{m}} = 
dX^{\bar{m}}/d\tau$, it can be checked easily that $G_{\bar{m}\bar{n}} 
u^{\bar{m}} u^{\bar{n}} = -1$. The momentum of the particle is obtained 
from \eqref{covariantparticleaction} to be $P_{\bar{m}} = \tau_{p}V_{p} 
e^{-\bar{\phi}} G_{\bar{m}\bar{n}} u^{\bar{n}}$, which leads to $P^{2} = 
G^{\bar{m}\bar{n}}P_{\bar{m}}P_{\bar{n}} = - (\tau_{p}V_{p})^{2}
e^{-2\bar{\phi}}$ 
and this makes it clear that such a particle is not tachyonic in the 
conventional sense unless the dilaton field becomes complex.

Therefore, to understand the origin of the tachyonic instability, 
let us analyze the equations of motion following from 
\eqref{covariantparticleaction} which correspond to the motion of a 
particle in a curved background subject to an acceleration, namely, 
\be
\frac{d^2 X^{\bar{m}}}{d\tau^2} + \Gamma_{\bar{n}\bar{p}}^{\bar{m}}\, \frac{d X^{\bar{n}}}{d\tau} \frac{d
  X^{\bar{p}}}{d\tau} = \alpha^{\bar{m}},\label{eqns}
\ee
where $\Gamma^{\bar{m}}_{\bar{n}\bar{p}}$ is the Christoffel symbol
constructed 
from the metric $G_{\bar{m}\bar{n}}$, namely,
\begin{equation}
\Gamma^{\bar{m}}_{\bar{n}\bar{p}} = \frac{1}{2}\,G^{\bar{m}\bar{q}}
\left(\partial_{\bar{n}}G_{\bar{q}\bar{p}} +
  \partial_{\bar{p}}G_{\bar{n}\bar{q}} 
- \partial_{\bar{q}}G_{\bar{n}\bar{p}}\right),\label{connection}
\end{equation}
and the proper acceleration $\alpha^{\bar{m}}$ is given by
\be
\alpha^{\bar{m}} = \left(G^{\bar{m}\bar{n}} + \frac{dX^{\bar{m}}}{d\tau}\frac{dX^{\bar{n}}}{d\tau}\right)
\partial_{\bar{n}}\bar{\phi}.\label{acceleration}
\ee
Thus we note that the dilaton background is responsible for a 
proper acceleration leading to a deviation of the trajectory of the
particle from its geodesic. It can be checked easily that  $G_{\bar{m}\bar{n}}
u^{\bar{m}} \alpha^{\bar{n}} = 0$ 
so that the proper acceleration is orthogonal to the proper velocity as would
be expected for a relativistic system. 
This is reminiscent of a Rindler particle executing hyperbolic motion \cite{Rindler}  and
clarifies the origin of the 
hyperbolic solution obtained in \cite{Kutasov:2004dj}. 

We can compute  the energy-momentum tensor by using
the general formula
\be
T^{\mu\nu} = - \frac{\partial L}{\partial(\partial_\mu X^m)}
\partial^\nu X^m + \eta^{\mu\nu} L,\label{tmunu}
\ee
where $L$ is the Lagrangian  of the action (5) and the non-vanishing 
components take the explicit forms,
\bea
T^{00} &=& \tau_p V_p \gamma e^{-\bar{\phi}} \equiv E,\nnu\\
T^{ij} &=& -\tau_p V_p \gamma^{-1} e^{-\bar{\phi}} \delta^{ij} \equiv p
\delta^{ij},\label{energypressure} 
\eea
where $\gamma$ is the Lorentz factor defined in \eqref{lorentzfactor} and $E, p$ denote the
energy and the pressure of the system respectively. From  time translation 
invariance we expect energy to be conserved and similarly rotational 
invariance in the transverse space leads to the conservation of angular momentum in the system. 

The time component ($\bar{m}=0$) of the equation of motion \eqref{eqns} yields
\be
\frac{d\gamma}{dt} = \gamma \frac{d\bar{\phi}}{dt},\label{energyconservation}
\ee
which we recognize from \eqref{energypressure} to lead to 
conservation of energy. On the other hand,  using 
 \eqref{energyconservation} the dynamical equation  \eqref{eqns} for $\bar{m}=m$  takes the form
\be
\ddot X^m + \Gamma^m_{\bar{n}\bar{p}} \dot X^{\bar{n}} \dot X^{\bar{p}} + G_{\bar{p}\bar{q}} \dot X^{\bar{p}} \dot
  X^{\bar{q}}  G^{mn} \partial_n\bar{\phi} = 0.\label{meqn}
\ee
We note that for large separations the leading behaviour of this equation is the free
particle motion described by $\ddot X^m = 0$. This corresponds to the
vanishing of the gravitational force as well as the  
acceleration $\alpha^m$ for large separations in the leading order. We study 
below the dynamics of the system in the next to leading order.   

For this purpose it is simpler to work in the
spherical-polar coordinates. Using the fact that angular momentum is 
conserved, we can restrict the motion of the particle to a
plane with the radial mode $R$ and the angular mode $\Theta$. In this case
the line element in \eqref{backgrounds} takes the form 
\be
- d\tau^2 = - dt^2 + H dR^2 + H R^2 d\Theta^2,\label{sphericallineelement}
\ee
and correspondingly the Lorentz factor \eqref{lorentzfactor} becomes,
\be
\gamma = \frac{1}{\sqrt{1-H\dot R^2 - H R^2 \dot \Theta^2}}.\label{sphericallorentzfactor}
\ee
The nonvanishing components of $\Gamma$ can be computed from \eqref{connection} and are given by,
\bea
\Gamma^R_{RR} &=& \partial_R \ln \sqrt{H},\quad 
\Gamma_{\Theta
  \Theta}^R = -R^{2} \partial_R\ln \sqrt{H R^2},\nnu\\
\Gamma^{\Theta}_{R \Theta} &=& \Gamma^{\Theta}_{\Theta R} =  \partial_R \ln
\sqrt{H R^2 }.\label{sphericalconnection}
\eea
In the spherical coordinates \eqref{meqn} has the form
\bea
\ddot R - R \dot \Theta^2 + \frac{1}{2 H^2} \partial_R H \left(2 H \dot R^2 -
  1\right) &=& 0,\nnu\\
\ddot \Theta + \frac{1}{H R^2} \partial_R \left(H R^2 \right) \dot \Theta \dot
R &=& 0.\label{sphericaleqns}
\eea
Since $\Theta$ is an angular coordinate, its conjugate gives the angular
momentum of the form  
\be
L = \tau_p V_p \gamma e^{-\bar{\phi}} H R^2 \dot \Theta = E H R^2 \dot \Theta.\label{angmom}
\ee
Defining the quantity
\be
H R^2 \dot \Theta = \frac{L}{E} = \ell,\label{ell}
\ee  
we note that the second equation in \eqref{sphericaleqns} leads to the conservation condition
\be
\frac{d}{dt} (H R^2 \dot \Theta) = \frac{d}{dt} \left(\frac{L}{E}\right) = \frac{d\ell}{dt}= 0,\label{angmomconservation}
\ee
namely,  the angular momentum associated with the motion of the particle is conserved. 

The true dynamics of the system is contained in the $R$-equation in
\eqref{sphericaleqns}. 
Using \eqref{angmom}  as well as the fact that energy in \eqref{energypressure} is conserved, 
we obtain from \eqref{sphericallorentzfactor}
\be
\dot{R}^{2} = \frac{1}{H}\Big(1 -
\frac{1}{H}\Big(\Big(\frac{\tau_{p}V_{p}}{E}\Big)^{2} + 
\frac{\ell^{2}}{R^{2}}\Big)\Big) \geq 0,\label{bounds}
\ee
which determines (using the form of $H$ in \eqref{backgrounds}) that for
$\frac{\tau_{p}V_{p}}{E}\geq 1$, 
we must have $(N\ell_{s}^{2}-\ell^{2})\geq 0$ and $R^{2} \leq R_{0}^{2}$,
while for 
$\frac{\tau_{p}V_{p}}{E}\leq 1$, we can have either
$(N\ell_{s}^{2}-\ell^{2})\geq 0$ without any restriction on $R$, 
or $(N\ell_{s}^{2} - \ell^{2})\leq 0$ with $R^{2} \geq R_{0}^{2}$ where
$R_{0}^{2} = 
\left|\frac{N\ell_{s}^{2}-\ell^{2}}{\big(\frac{\tau_{p}V_{p}}{E}\big)^{2}-1}\right|$. Since 
we are interested in the behavior of the system close to the origin $R\simeq 
0$, 
it is clear that we must have $(N\ell_{s}^{2} - \ell^{2})\geq 0$ independent 
of the value of the ratio $\frac{\tau_{p}V_{p}}{E}$. (Parenthetically, we
remark 
here that the analysis in \cite{Kutasov:2004dj} assumed that 
$\frac{\tau_{p}V_{p}}{E}\geq 1$, but the difference in whether this 
ratio is bigger than or smaller than unity simply reflects how far 
away the D$p$-brane can be from the NS5 branes initially. Since it 
is natural to assume that the D$p$-brane starts out infinitely far 
away, we would assume $\frac{\tau_{p}V_{p}}{E}\leq 1$, although what 
is really important for the analysis of the behavior near the NS5 
branes is that $N\ell_{s}^{2}-\ell^{2}\geq 0$.)

From \eqref{sphericalconnection} as well as \eqref{acceleration} we 
find that for large $R$ both the gravitational  as well as the dilatonic 
forces behave as $\frac{1}{R^{3}}$ so that  for large $R \gg \sqrt{N}\ell_{s}$ 
the radial equation in \eqref{sphericaleqns} yields 
\be
\ddot R + (N\ell_{s}^{2} - \ell^{2})\left(-\frac{2N\ell_{s}^{2}}
{R_{0}^{2}} +1\right)\frac{1}{R^3}  = 0,\label{largeReqn}
\ee
as expected for a particle moving in an attractive $-1/R^2$ potential 
(up to multiplicative factors) in
four spatial dimensions (as long as $(N\ell_{s}^{2}-\ell^{2})>0$). 
The dynamics of the D$p$-brane in the 
vicinity ($R\ll\sqrt{N}\ell_{s}$) of the NS5-branes is better 
understood in the variable $Z=1/R$. In this variable
the harmonic function becomes $H = 1 + N \ell_s^2 Z^2$ and 
for $\sqrt{N} \ell_s Z \gg 1$, the $R$ equation in \eqref{sphericaleqns} takes the form
\be
\ddot Z - \frac{1}{(N\ell_{s}^{2})^{2}} (N \ell_s^2 - \ell^2) Z
= 0.\label{smallReqn}
\ee
For $(N\ell_{s}^{2}-\ell^{2})> 0$ which is the case of interest for us, 
we recognize \eqref{smallReqn} to correspond to an inverted simple harmonic 
oscillator. On the other hand, we note that in the absence of the acceleration 
the radial equation for small $R$ reduces to $\ddot Z = 0$.  
The origin of the tachyonic instability is now clear, namely, it is the
acceleration due to the dilatonic background which is the source 
of the instability. Even though $P^{2}< 0$ indicating that the particle is not tachyonic in 
the conventional sense, the dilatonic force that it experiences in the 
background of the NS5 branes leads to hyperbolic motion 
and  the tachyonic instability in the system. Furthermore, using this hyperbolic
motion (solution of \eqref{smallReqn}), we can check the known fact
that the pressure as given in \eqref{energypressure} falls off 
exponentially at late time. We note here that although  
\eqref{smallReqn} exhibits a tachyonic instability,  in order to
see where the instability occurs we give the complete radial equation below
(see the first eq. in \eqref{sphericaleqns}) 
\bea
\ddot R &=& \frac{(N \ell_s^2 - \ell^2) N\ell_s^2}{(R^2 + N \ell_s^2)^3}
  \left(R + \left(\frac{2}{R_0^2}-\frac{1}{N\ell_s^2}\right)R^3\right)\nnu\\ 
&=& -\frac{d V}{d R}. \label{smallReqnnextorder}
\eea
where $V(R)$ is the potential in which the particle moves.
From this we find that, for $R_0 < \sqrt{N} \ell_s$ which is
the case of our interest, the instability occurs at $R=0$ and
the effective mass squared of the particle is  
\be
m^{2} \equiv \left.\frac{d^2 V(R)}{dR^2}\right|_{R=0} = - \frac{(N\ell_{s}^{2} - 
\ell^{2})}{(N\ell_{s}^{2})^{2}}.\label{tachyonmass}
\ee
However, since the effective string coupling $e^{\bar\phi}$ blows up at $R=0$,
a full quantum treatment is necessary to identify the true position of instability.  

We emphasize that 
in the absence of the additional force produced by the dilaton, the particle 
will follow a gravitational geodesic which does not lead to any
instability. Of course,  this cannot happen within the context of string theory 
unless we set the dilaton to zero by hand. On the other hand, if we consider
the motion of a fundamental string, given by the Nambu-Goto action, 
in the background of a stack of coincident D5-branes, then the absence of a  
dilaton prefactor in the Nambu-Goto action may lead one to think that the 
particle will not experience any acceleration and would follow the 
gravitational geodesic without any
instability. However, this is not true since the background D5-branes can 
give rise to an induced metric with an overall 
conformal factor (so that the action takes the form \eqref{covariantparticleaction}) 
and this conformal factor can indeed be a source of  proper acceleration 
leading to a geometric tachyon in the dynamics. In fact, it is known 
that a parallel but separated F-D5 system is non-supersymmetric and
so we expect a tachyonic instability.
The F-string in this case is known to melt into the D5-branes to 
form a non-threshold bound state \cite{Lu:1999uca}. The formation of this 
bound state can be viewed as due to the geometric tachyon condensation. 
The details of this work will be published in a companion paper \cite{Das}.

This work was supported in part by US DOE Grant number DE-FG 02-91ER40685.
Two of us (S.P. and S.R.) would like to thank the Department of Physics and
Astronomy of the University of Rochester, where this work is done, for the
warm hospitality.

\end{document}